\documentstyle[11pt,amssymb,epsf]{article}

\def\ben{\begin{equation}}
\def\een{\end{equation}}

  \let\n=\nu  \let\p=\pi

\let\C=\Chi

\def\nn{\nonumber} \def\bd{\begin{document}} \def\ed{\end{document}}
\def\ds{\documentstyle} \let\fr=\frac \let\bl=\bigl \let\br=\bigr
\let\Br=\Bigr \let\Bl=\Bigl
\let\bm=\bibitem
\let\na=\nabla
\let\pa=\partial \let\ov=\overline
\newcommand{\be}{\begin{equation}}
\newcommand{\ee}{\end{equation}}
\def\ba{\begin{array}}
\def\ea{\end{array}}
\def\ft#1#2{{\textstyle{{\scriptstyle #1}\over {\scriptstyle #2}}}}
\def\fft#1#2{{#1 \over #2}}
\def\del{\partial}
\def\vp{\varphi}
\def\sst#1{{\scriptscriptstyle #1}}
\def\oneone{\rlap 1\mkern4mu{\rm l}}
\def\td{\tilde}
\def\wtd{\widetilde}
\def\ie{\rm i.e.\ }
\def\dalemb#1#2{{\vbox{\hrule height .#2pt
        \hbox{\vrule width.#2pt height#1pt \kern#1pt
                \vrule width.#2pt}
        \hrule height.#2pt}}}
\def\square{\mathord{\dalemb{6.8}{7}\hbox{\hskip1pt}}}
\newcommand{\ho}[1]{$\, ^{#1}$}
\newcommand{\hoch}[1]{$\, ^{#1}$}
\newcommand{\bea}{\begin{eqnarray}}
\newcommand{\eea}{\end{eqnarray}}
\newcommand{\ra}{\rightarrow}
\newcommand{\lra}{\longrightarrow}
\newcommand{\Lra}{\Leftrightarrow}
\newcommand{\bp}{\tilde \beta^\prime}
\newcommand{\tr}{{\rm tr} }
\newcommand{\Tr}{{\rm Tr} }
\def\0{{\sst{(0)}}}
\def\1{{\sst{(1)}}}
\def\2{{\sst{(2)}}}
\def\3{{\sst{(3)}}}
\def\4{{\sst{(4)}}}
\def\5{{\sst{(5)}}}
\def\6{{\sst{(6)}}}
\def\7{{\sst{(7)}}}
\def\8{{\sst{(8)}}}
\def\n{{\sst{(n)}}}
\def\cA{{{\cal A}}}
\def\cB{{{\cal B}}}
\def\cF{{{\cal F}}}
\def\cH{{{\cal H}}}
\def\tV{\widetilde V}
\def\tW{\widetilde W}
\def\tH{\widetilde H}
\def\tE{\widetilde E}
\def\tF{\widetilde F}
\def\tA{\widetilde A}
\def\im{{i}}
\def\tY{{{\wtd Y}}}
\def\ep{{\epsilon}}
\def\vep{{\varepsilon}}
\def\R{\rlap{\rm I}\mkern3mu{\rm R}}
\def\bD{{{\bar D}}}

\def\R{\rlap{\rm I}\mkern3mu{\rm R}}
\def\bD{{{\bar D}}}
\def\R{{{\Bbb R}}}
\def\C{{{\Bbb C}}}
\def\H{{{\Bbb H}}}
\def\CP{{{\Bbb C}{\Bbb P}}}
\def\RP{{{\Bbb R}{\Bbb P}}}
\def\Z{{{\Bbb Z}}}
\def\bA{{{\Bbb A}}}
\def\bB{{{\Bbb B}}}
\def\bC{{{\Bbb C}}}
\def\bD{{{\Bbb D}}}
\def\bE{{{\Bbb E}}}
\def\bZ{{{\Bbb Z}}}
\def\Re{{{\frak{Re}}}}
\def\Im{{{\frak{Im}}}}
\def\cosec{{\,\hbox{cosec}\,}}
\def\Gm{{\Gamma_{\!\! -}}}
\def\Gp{{\Gamma_{\!\! +}}}
\def\stan{{standard }}
\def\nonstan{{supernumerary }}
\def\p{{\partial}}

\def\bog{{Bogomol'nyi\ }}

\newcommand{\tamphys}{\it Center for Theoretical Physics,
Texas A\&M University, College Station, TX 77843}

\newcommand{\upenn}{\it Department of Physics and Astronomy,\\ University
of Pennsylvania, Philadelphia, PA 19104}

\newcommand{\brussels}{\it Physique Th\'eorique et Math\'ematique,
Universit\'e Libre de Bruxelles,\\ Campus Plaine C.P. 231, B-1050
Bruxelles, Belgium}

\newcommand{\damtp}{\it DAMTP, Centre for Mathematical Sciences,\\
 Cambridge University, Wilberforce Road, Cambridge CB3 OWA, UK}

\newcommand{\auth}{Z.W. Chong\hoch{\ddagger}, M. Cveti\v c\hoch{*2},   
H. L\"u\hoch{\ddagger1} and C.N. Pope\hoch{\ddagger1}}

\begin{document}

\begin{flushright}
MIFP-05-11\ \ \ UPR-1123-T \\
{\bf hep-th/0505112}\\
May\  2005
\end{flushright}

\vspace{10pt}

\begin{center}

{\large {\bf Five-Dimensional Gauged Supergravity Black Holes with
 Independent Rotation Parameters}}

\vspace{20pt}
\auth

\vspace{10pt}{\hoch{*}\it Department of Physics and Astronomy,\\
University of Pennsylvania, Philadelphia, PA 19104, USA}

\vspace{10pt}{\hoch{\ddagger}\it George P. \& Cynthia W. Mitchell
Institute for Fundamental Physics,\\ Texas A\& M University,
College Station, TX 77843-4242, USA}


%
%
%

\vspace{20pt}


\begin{abstract}

    We construct new non-extremal rotating black hole solutions in
$SO(6)$ gauged five-dimensional supergravity.  Our solutions are the
first such examples in which the two rotation parameters are
independently specifiable, rather than being set equal.  The black
holes carry charges for all three of the gauge fields in the $U(1)^3$
subgroup of $SO(6)$, albeit with only one independent charge
parameter.  We discuss the BPS limits, showing in particular that
these include the first examples of regular supersymmetric black holes
with independent angular momenta in gauged supergravity.  We also find
non-singular BPS solitons.  Finally, we obtain another independent
class of new rotating non-extremal black hole solutions with just one
non-vanishing rotation parameter, and one non-vanishing charge.

\end{abstract}
\end{center}

{\vfill\leftline{}\vfill \vskip 10pt \footnoterule {\footnotesize
{\footnotesize
\hoch{1} Research supported in part by DOE grant
DE-FG03-95ER40917.}\vskip 2pt
\hoch{2} Research supported in part by DOE grant
DE-FG02-95ER40893, NSF grant INTO3-24081, and the\\
$\phantom{xxxxi}$  Fay R. and Eugene L.
Langberg Chair.}\vskip 2pt
}

\pagebreak

\newpage

\section{Introduction}

    Constructing non-extremal charged rotating black hole solutions in 
gauged supergravity is quite a complicated problem. This is because, 
unlike the case of ungauged supergravity, there are no known 
solution-generating techniques that could be used to add charges to the 
the already-known neutral rotating black hole solutions found in
four dimensions in \cite{carter}, five dimensions in \cite{hawhuntay}, 
and $D\ge6$ dimensions in \cite{gilupapo1,gilupapo2}. Aside from the
four-dimensional Kerr-Newman-AdS black holes, which were found in
\cite{carter2}, the known non-extremal charged rotating black hole 
solutions comprise recently-discovered examples in five-dimensional
gauged supergravities in \cite{d5gauge1,d5gauge2}; in four-dimensional
gauged supergravity in \cite{d4gauge}; and in seven-dimensional 
gauged supergravity in \cite{d7gauge}.  In the five and seven-dimensional
cases, the problem was simplified greatly by taking the {\it a priori}
independent rotation parameters of the orthogonal 2-planes in the
transverse space to be equal.  This reduces the problem to studying
cohomogeneity-1 metrics, with non-trivial coordinate dependence on only
the radial variable, rather then metrics of cohomogeneity 2 or 
cohomogeneity 3.

   In this paper, we shall present some new results on non-extremal
rotating black holes in five-dimensional gauged supergravity, in which
the two rotation parameters $a$ and $b$ can be independently
specified.  Our black holes can be viewed as solutions in $N=8$ gauged
$SO(6)$ supergravity, with three charge parameters associated with the
gauge fields of the $U(1)^3$ abelian subgroup.  They can also be
viewed as solutions in $N=2$ gauged supergravity coupled to two vector
multiplets.  Our solutions are not the most general possible, in that
there is a specific relation between the three charges.  However, the
metrics have considerable interest because they allow us to study the
thermodynamics, and the supersymmetric limits, for a rather general
class of non-extremal black holes with unequal angular momenta.

   After presenting the solutions, we then calculate the charges, 
angular momenta, angular velocities, electrostatic potentials, 
temperature and entropy.  From these, we follow the procedure 
that was used in \cite{gibperpop}, and more recently in \cite{cvgilupo},
for calculating the energy by integration of the first law of 
thermodynamics.  The fact that this is possible at all, \ie that
the right-hand side of the first law is an exact differential, is already a 
highly non-trivial test of the validity of the thermodynamic relations
for these black holes.

   Having obtained expressions for the energy $E$, the angular momenta
$J_a$ and $J_b$, and the three charges $Q_i$, we can study the
conditions under which supersymmetric limits will arise, by looking
for zero eigenvalues of the \bog matrix arising from anticommutators
of the supercharges.  We obtain by this means families of
supersymmetric configurations, characterised by a mass parameter and
the two independent rotation parameters.  In general these BPS
solutions have naked closed timelike curves (CTC's) lying outside a
Killing horizon.  However, for a particular choice of the mass, we
obtain completely regular black holes with no singularities or closed
timelike curves on or outside the horizon.  These are similar to the
regular black holes of five-dimensional gauged supergravity that were
found in \cite{gutrea1,gutrea2}, except that in our new solutions the
two angular momenta can be independently specified. Indeed, the
rotating BPS black holes that we find in this paper are the first such
examples with independent rotation parameters.  We also find other
special cases, describing completely regular solitons.

   We also obtain a further class of new non-extremal rotating black
hole solutions of five-dimensional gauged supergravity, in which only
one rotation parameter is non-vanishing, and only one of the three
$U(1)$ charges is turned on.  These solutions are therefore independent
of any found previously in this paper or elsewhere.  We again study the
thermodynamics and obtain expressions for the conserved energy, angular
momentum and charge.  From the BPS limit we again obtain supersymmetric 
solutions.  In this case, unlike the one discussed above, there are
no regular BPS black holes or solitons, but only solutions with naked CTC's. 

\section{Black Holes with Two Unequal Rotations}\label{3absec}

   The rotating black hole metrics that we shall construct arise as 
solutions of $SO(6)$ gauged five-dimensional supergravity.  They are 
charged under the $U(1)^3$ Cartan subgroup of $SO(6)$, with specific
relations between the three charges.  The two rotation parameters can be
specified independently.  The relevant part
of the supergravity Lagrangian that describes these solutions is given by
\be
e^{-1}\, {\cal L} = R - \ft12{\del\vec\varphi}^2 -
  \ft14\sum_{i=1}^3 X_i^{-2}\, {(F^i)}^2  + 4 g^2 \,
  \sum_{i=1}^3 X_i^{-1} + \ft1{24} \ep_{ijk}\, \ep^{\mu\nu\rho\sigma\lambda}
  F^i_{\mu\nu}\, F^j_{\rho\sigma}\, A^k_{\lambda}\,,\label{d5lag}
\ee
where $\vec\varphi=(\varphi_1,\varphi_2)$, and
\be
X_1= e^{-\fft1{\sqrt6}\varphi_1 -\fft1{\sqrt2} \varphi_2}\,,\qquad
X_2= e^{-\fft1{\sqrt6}\varphi_1 +\fft1{\sqrt2} \varphi_2}\,,\qquad
X_3 = e^{\fft2{\sqrt6}\varphi_1}\,.
\ee

   In the following subsections we shall first construct the
non-extremal rotating metrics, and then calculate their associated 
conserved quantities, namely their mass $E$, their angular momenta $J_a$
and $J_b$, and the three electric charges $Q_i$.  Next, by using the
BPS conditions derived from the AdS superalgebra, we determine the
restrictions on the parameters of the solutions that lead to
supersymmetry.  We investigate the global structure of these BPS
limits, showing in particular that there exist regular supersymmetric
black holes with no naked singularities or closed timelike curves.

\subsection{The Non-Extremal Black Holes}\label{abbhsec}

   Since there are no solution-generating techniques available for 
constructing non-extremal rotating black holes in gauged supergravities,
our procedure for obtaining them depends to a large extent on a 
combination of guesswork and conjecture, followed by an explicit verification
that the equations of motion are indeed satisfied.  Here, we simply present
the outcome of this process.

   We find that the following provides a solution of the five-dimensional 
gauged supergravity equations:
\bea
ds^2&=&\,H^{-\fft43}\,\Big [ -\fft{X}{\rho^2}\,
  (dt-a\,\sin^2\theta\,\fft{d\phi}{\Xi_a}-
   b\,\cos^2\theta\,\fft{d\psi}{\Xi_b})^2\nn\\
&&+\fft{C}{\rho^2}(\,\fft{ab}{f_3}dt-\fft{b}{f_2}
   \sin^2\theta \fft{d\phi}{\Xi_a}-
   \fft{a}{f_1}\cos^2\theta \fft{d\psi}{\Xi_b})^2\nn\\
&+&\fft{Z\sin^2\theta}{\rho^2}(\fft{a}{f_3}dt-
   \fft{1}{f_2}\fft{d\phi}{\Xi_a})^2+
  \fft{W\cos^2\theta}{\rho^2}(\fft{b}{f_3}dt-
   \fft{1}{f_1}\fft{d\psi}{\Xi_b})^2\Big ]+
H^\fft23\,(\fft{\rho^2}{X}dr^2+\fft{\rho^2}{\Delta_\theta}d\theta^2\,)\,,
\nn\\
H&=&\tilde \rho^2/\rho^2,\quad \rho^2=r^2\,+\,a^2\,\cos^2\theta\,
   +\,b^2\,\sin^2\theta,\quad \tilde \rho^2=\rho^2\,+\,2\,m\,s^2\,,\nn\\
f_1&=&a^2\,+\,r^2,\quad f_2=b^2\,+\,r^2,\quad f_3=(a^2+r^2)(b^2+r^2)\,+
   \,2\,m\,r^2\,s^2;\nn\\
\Delta_\theta &=&1-a^2\,g^2\,\cos^2\theta-b^2\,g^2\,\sin^2\theta,\nn\\
X&=&\fft{1}{r^2}(a^2+r^2)(b^2+r^2)-2m+g^2(a^2+r^2+2ms^2)(b^2+r^2+2ms^2)
 \,,\label{d5sol}\\
C&=&f_1\,f_2(X\,+\,2m\,-\,4\,m^2\,s^4/\rho^2),\nn\\
Z&=&-b^2\,C\,+\,\fft{f_2\,f_3}{r^2}[ f_3\,-\,g^2\,r^2\,
             (a^2\,-b^2)(a^2+r^2+2\,m\,s^2)\cos^2\theta],\nn\\
W&=&-a^2\,C\,+\,\fft{f_1\,f_3}{r^2}[ f_3\,+
  \,g^2\,r^2\,(a^2\,-b^2)(b^2+r^2+2\,m\,s^2)\sin^2\theta]\,,\nn\\
\Xi_a &=& 1 -a^2 g^2\,,\qquad \Xi_b= 1 - b^2 g^2\,,\nn
\eea
where
\be
s\equiv \sinh\delta\,,\qquad c\equiv \cosh\delta\,.
\ee

   The gauge potentials and scalar fields are given by
\bea
A^1&=&A^2=\fft{2\,m\,s\,c}{\tilde \rho^2}(dt\,-\,a\,\sin^2\theta\,
  \fft{d\phi}{\Xi_a}\,-\,b\cos^2\,\theta\,\fft{d\psi}{\Xi_b})\nn\\
A^3&=&\fft{2\,m \,s^2}{\rho^2}(b\,\sin^2\theta\,\fft{d\phi}{\Xi_a}\,+\,a\,
  \cos^2\theta\,\fft{d\psi}{\Xi_b})\nn\\
X_1&=&X_2=H^{-\fft13},\quad X_3=H^{\fft23}\,.\label{axsol}
\eea

  It should be noted that the solution above is presented in a 
coordinate frame that is rotating at infinity.  One can pass to coordinates
that are asymptotically static by making the redefinitions
\be
\phi = \td\phi + a g^2\, t\,,\qquad
\psi=\td\psi + b g^2\, t\,.
\ee
It is helpful to make this transformation in order to simplify the
calculation of the thermodynamic quantities.  One might think from the
expressions for the gauge potentials in (\ref{axsol}) that there
are just two non-vanishing (and equal) charges, since $A^3$ has no
electric component.  However, this is a somewhat misleading artefact
of the original rotating coordinate system.  After transforming to
the asymptotically non-rotating frame, one finds that $A^3$ also
has an electric component, and indeed, as we shall see below, the
third electric charge is non-zero too.

   It is straightforward to calculate the temperature, entropy,
angular velocities on the horizon, and the electrostatic potentials
on the horizon, referred to the asymptotically static frame.  We find
\bea
T &=& \fft{2 g^2\, r_+^6 + [1+g^2 (a^2+b^2+ 4 m s^2)] r_+^4 -a^2 b^2}{
        2\pi r_+[r_+^4 + (a^2+b^2+2m s^2) r_+^2 + a^2 b^2}\,,\nn\\
S &=&\fft{\pi^2 [r_+^4 + (a^2+b^2+2m s^2) r_+^2 + a^2 b^2]}{
2 \Xi_a\, \Xi_b\, r_+}\,,\nn\\
\Omega_a &=& \fft{a\, [g^2 r_+^4 + (b^2+2ms^2) g^2 r_+^2 + b^2]}{
       r_+^4 + (a^2+b^2+2m s^2) r_+^2 + a^2 b^2}\,,\nn\\
\Omega_b &=& \fft{b\, [g^2 r_+^4 + (a^2+2ms^2) g^2 r_+^2 + a^2]}{
       r_+^4 + (a^2+b^2+2m s^2) r_+^2 + a^2 b^2}\,,\nn\\
\Phi_1&=&\Phi_2 = \fft{2m r_+^2\, s c}{r_+^4 
            + (a^2+b^2+2m s^2) r_+^2 + a^2 b^2}\,,\nn\\
\Phi_3 &=& \fft{2mab s^2}{r_+^4 + (a^2+b^2+2m s^2) r_+^2 + a^2 b^2}\,,
\eea
where $r_+$, the largest root of the metric function $X(r)$, is the 
location of the outer horizon.  

   The two angular momenta can be evaluated from Komar integrals
\be
J = \fft1{16\pi}\, \int_{S^3} {*dK}\,,
\ee
where $K$ is the 1-form obtained by lowering the index on the angular
Killing vector $\del/\del\phi$ or $\del/\del\psi$.  The charges are given
by Gaussian integrals 
\be
Q_i = \fft1{16\pi}\, \int_{S^3} (X_i^{-2} {*F^i} -\ft12 \epsilon_{ijk}\, 
       A^j\wedge A^k)\,,
\ee
Having evaluated the angular momenta and charges, we can then integrate the
first law of thermodynamics
\be
dE = TdS + \Omega_a\, dJ_a + \Omega_b\, dJ_b + \sum_i \Phi_i\, dQ_i
\ee
in order to obtain the energy $E$ of the black hole solution.  Our results
for the conserved quantities are
\bea
E&=& \fft{\pi m[2\Xi_a + 2\Xi_b - \Xi_a\, \Xi_b +
                  (2\Xi_a^2 + 2\Xi_b^2 + 2\Xi_a\, \Xi_b 
      - \Xi_a^2\, \Xi_b - \Xi_b^2\, \Xi_a)\, s^2 ]}{4 \Xi_a^2\, \Xi_b^2}\,,\\
J_a &=& \fft{\pi m a\,  (1 + s^2\, \Xi_b)}{2 \Xi_b\, \Xi_a^2}\,,\qquad
J_b = \fft{\pi m b\,  (1 +  s^2 \, \Xi_a)}{2 \Xi_a\, \Xi_b^2}\,,\nn\\
Q_1&=& Q_2= \fft{\pi m s c}{2 \Xi_a\, \Xi_b}\,,\qquad
Q_3 = -\fft{\pi a b m s^2  g^2}{2 \Xi_a\, \Xi_b}\,.\label{thermo1}
\eea

\subsection{The BPS limit}\label{abbpssec}

   As discussed recently in \cite{cvgilupo}, the BPS limit of the 
non-extremal solution can conveniently be discussed by studying the
eigenvalues of the \bog matrix that arises from the anticommutator
of the supercharges of the AdS superalgebra.  Thus 
the BPS limit of non-extremal
five-dimensional gauged supergravity solutions is attained 
when\footnote{The \bog matrix actually has four eigenvalues,
namely $E\pm g J_a \pm g J_b -\sum_i Q_i$.  They are equivalent, under
reversal of the signs of the rotation parameters, and so without
loss of generality we have chosen just to consider one of them.} 
\be
E+ g J_a + g J_b - \sum_i Q_i =0\,.
\ee
Substituting our expressions for the conserved mass, angular momenta
and charges given in (\ref{thermo1}), we find that the BPS condition
is satisfied if the parameter $\delta$ is chosen so that
\be
e^{2\delta} = 1 + \fft{2}{(a+b)\, g}\,.\label{bps1}
\ee
In section (\ref{abglobsec}), we shall discuss the global structure
of the BPS solutions.
   
    It is interesting to note that with $a$ and $b$ as independently
specificable parameters, we can make contact with previous results in
two inequivalent special cases.  Firstly, if we take $a=b$, the
solutions we have obtained in this paper reduce to particular cases of
the 3-charge rotating black holes with equal angular momenta that were
found in \cite{d5gauge2}.  In particular, the BPS condition
(\ref{bps1}) reduces to one that was found for the $a=b$ solutions in
\cite{cvgilupo}.  An inequivalent special case arises if instead we
take $a=-b$.  Now, the BPS condition (\ref{bps1}) reduces to the
condition that $e^{2\delta}\rightarrow \infty$, which was also arose,
as a disjoint case, in the analysis in \cite{cvgilupo}; it again can
be viewed as a situation with ``equal angular momenta,'' after making
an orientation reversal.  Because in the present work we have the
possibility to specify $a$ and $b$ independently, we can actually
describe a continuous interpolation between two BPS limits that were
seen as disjoint possibilities in the earlier work.

\subsection{Global analysis}\label{abglobsec}

       To analyse the global structure of the metric (\ref{d5sol}), we
first rewrite it in the form
\bea
ds^2&=&H^{2/3}\Big[ - \fft{X\, \Delta_\theta\, \sin^2 2\theta}{4
\Xi_a\Xi_b\, H^2 B_\phi B_\psi}\, r^2dt^2 + \rho^2(
\fft{dr^2}{X} +\fft{ d\theta^2}{\Delta_\theta}) +
B_\psi (d\psi + v_1 d\phi + v_2 dt)^2\nn\\
&& \qquad\qquad + B_\phi (d\phi + v_3 dt)^2\Big]\,.
\eea
From this, it is evident that there is an outer Killing horizon located at 
$r=r_+$, the largest root of $X(r)$.  There is a velocity of
light surface (VLS), located at the boundary $r=r_L$ of the
region where $B_\psi B_\phi$ changes sign from positive (at large $r$) to
negative.  Inside
the VLS, the metric develops closed timelike curves (CTC's).  
If $r_+ > r_L$, then the Killing horizon lies outside the VLS,
and so the Killing horizon is an event horizon. In these circumstances,
the solution describes a regular black hole, in which there are neither
curvature singularities nor CTC's outside the horizon.  If the largest root
$r_+$ is inside the VLS, the solution instead describes a naked time machine.

        In the supersymmetric limit, there exists a Killing vector
\be
\ell = \fft{\del}{\del t} - g\, \fft{\del}{\del\td\phi} -
 g \fft{\del}{\del \td\psi}
\ee
that has a spinorial square root, in the sense that 
$\ell \sim \bar\eta \gamma^\mu\eta\, \del_\mu$, where $\eta$ is the Killing
spinor.  (See \cite{cvgilupo} for a recent discussion of this.)  This
Killing vector is necessarily non-spacelike, and in fact we find that
the explicit expression for its norm is a manifestly non-positive quantity.
From this, we find the identity
\bea
\!\!\!&&-\fft{X\, \Delta_\theta \sin^2 2\theta}{4\Xi_a\Xi_b\,
H^2 B_\psi B_\phi}
+B_\psi (  v_2 + g - b g^2 + v_1 (g - a g^2) )^2 +
        B_\phi ( v_3 + g - a g^2 )^2\label{identity1} \\
\!\!\! &&= - \fft{\Big[(1+ag +bg)(1\! +\! a g \cos^2\theta\! 
+\! bg\sin^2\theta)
\! -\! H (ag(1+ ag) \cos^2\theta\! +\! 
      bg(1+bg)\sin^2\theta)\Big]^2}{(1+ag)^2(1+bg)^2 H^2}\nn
\eea

   It follows from (\ref{identity1}) that in general we shall have
$B_\phi B_\psi <0$ at the largest root $r=r_+$ where $X(r)$ vanishes.
This means that the VLS at $r=r_L$ lies outside the Killing horizon at
$r=r_+$, thus implying naked CTC's.  

     As in the cases discussed in \cite{cvgilupo}, there are two ways
of avoiding such naked CTC's.  The first is if the parameters are
chosen so that right-hand side of (\ref{identity1}) vanishes at 
$r=r_+ \equiv r_0$. This occurs if
\be
m=\fft{(a+b)^2(1+ag)(1+bg)(2+ ag + bg)}{2(1 + ag + bg)}\,.
\ee
Note that when this condition is satisfied, the function $X$ becomes
\be
X=\fft{(r-r_0)^2 ((ab-r_0)^2 r^2 + a^2b^2(a+b)^2)}{(a+b)^2r_0^4 r^2}\,,
\ee
and there is in fact a double root at $r=r_0$, where $r_0$ is given by
\be
r_0^2 = \fft{ab}{1 + a g + b g}\,.
\ee
Now it is straightforward to verify that the VLS lies inside the horizon at
$r=r_0$, and so the Killing horizon is an event horizon.  Thus the
solution describes a supersymmetric black hole that is regular on and
outside the event-horizon. The fact that $X(r)$ has a double root at
$r=r_0$ implies that the Hawking temperature is zero.  This is the
first example of a supersymmetric black hole with two independent
angular momenta in gauged supergravity.  If $a$ and $b$ are set equal, the
solution reduces to a special case of the regular black holes found in
\cite{gutrea2}.\footnote{Note that a black
hole is not possible if $a+b=0$.  As we discussed in section
\ref{abbpssec}, setting $a=-b$ corresponds to a BPS limit that was
also studied in \cite{cvgilupo}, which was associated with BPS
solutions found in \cite{klemm1,klemm2}.  The solutions in
\cite{klemm1,klemm2} all describe configurations with naked CTC's, or,
as shown in \cite{cvgilupo}, a non-singular soliton; there is no
solution in that family that describes a regular black hole.}  The
rotation parameters $a$ and $b$ must in general be restricted to an appropriate
range, in order to ensure that $B_\phi$ and $B_\psi$ remain positive and
hence that there are no CTC's outside the horizon.

   The other way to avoid the naked CTC's of the generic supersymmetric 
solutions is by restricting the parameters so that $B_\psi B_\phi$ goes to
zero at the same radius as $X$ goes to zero, \ie so that $r_L=r_+$.  
This occurs when the parameter $m$ is given by
\be
m=\fft{(a+b)(1+ag)(1+bg)(2+ag+bg)(1 + 2ag+bg)(1+ag+2bg)}{
2g (1 + ag + bg)^2}\,.
\ee
The solution then describes a smooth topological soliton.  Defining $R=r^2 +
a^2 b^2 g^2/(1 + ag + bg)^2$, the coordinate $R$ runs from 0 to $\infty$.
The requirement of having no conical singularity at $R=0$, where
$B_\phi=0$, implies the quantisation condition
\be
\fft{1 + 3(a+b)g+ (3a^2 + 5ab+3b^2)g^2 + (a+b)(a^2 + b^2)g^3 -
          ab(a^2 + 4ab+b^2) ^4}{bg(1-ag)(1+ag+bg)(1+2ag+bg)}=1\,.
\ee
In the special cases $a=b$ or $a=-b$, these topological solitons are
encompassed within the soliton solutions obtained in \cite{cvgilupo}.
The rotation parameters $a$ and $b$ must in general be restricted to
an appropriate range, in order to ensure that $B_\phi$ and $B_\psi$
remain positive and hence that there are no CTC's for all $R\ge 0$.

        Aside from the above two possibilities, the supersymmetric
solutions have naked CTC's in general.  As in the examples in
\cite{cvgilupo}, a conical singularity at the Killing horizon can be
avoided by periodically identify the asymptotic time coordinate $t$
with an appropriate period.  However, if the Killing horizon is
associated with a double root of $X(r)$, then such an identification
is unnecessary.

\section{Black Holes with One Rotation and One Charge}

   In this section we obtain another new solution describing a non-extremal
rotating black hole in gauged five-dimensional supergravity.  In this
case, just one of the two rotation parameters is non-zero, and only one of the
three gauge fields in the $U(1)^3$ subgroup of $SO(6)$ is turned on.  This
solution is therefore not a special case of the solution obtained in
section \ref{3absec}, nor indeed of any other previously-obtained 
solutions.  Having presented the solution, we then evaluate the conserved
mass, angular momentum and charge, and from this we study the BPS limit.

\subsection{The non-extremal solution}

    Again, since the process that led us to the solution was a long one,
involving making a conjecture for its form, and then verifying explicitly
that it solved the equations, we shall just present our final result here.
We find that the metric for the five-dimensional black hole
with one non-vanishing rotation parameter and one charge is given by 
\bea 
ds^2\,&=& -H^{-\fft23}\,\fft{w\,Y}{F(r,\theta)}\,(\,c\,dt\,-\,
a\,\sin^2\theta \fft{d\phi}{w\,\Xi})^2\nn\\
&+&H^{\fft13}\Big (\fft{w\,\Delta_\theta\sin^2\theta}{F(r,\theta)}
(\,f_1\, dt-c\,f_2\,\fft{d\phi}{w\,\Xi})^2+\fft{\rho^2}{Y}dr^2+
\fft{\rho^2}{\Delta_\theta}d\theta^2+r^2\cos^2\theta d\psi^2 \Big ),
\nn\\
F(r,\theta)\,&=&\,c^2\,f_2(r)\,-\,a\,\sin^2\theta\,f_1(r),\nn\\
H\,&=&\,\tilde\rho^2/\rho^2,\quad \tilde\rho^2\,=\,a^2\,+\,r^2\,+\,
2\,m\,s^2,\quad \rho^2\,=\,a^2\,+\,r^2,\nn\\
f_1&=& \fft{a}{w}\,(1-w\,s^2\,g^2\,(a^2+r^2)),
\quad f_2=\,w\,(a^2+r^2),\nn\\
Y&=&(a^2+r^2)(1+g^2\,r^2)\,-\,2\,m\,f_1\,/\,a,\nn\\
\Delta_\theta\,&=&\,1\,-\,a^2\,g^2\,\cos^2\theta,\quad 
\Xi\,=\,1-a^2\,g^2.
\eea
where again $c=\cosh\delta$, $s=\sinh\delta$, and the constant $w$,
which satisfies $c^2 w^2 - s^2 w \Xi =1$, is given by
\be
w=\fft{\Xi\, s^2 + \sqrt{4(1 + s^2) + \Xi^2\,  s^4}}{2(1+s^2)}
\,.
\ee
The gauge potentials and scalar fields are given by
\bea
A^1\,&=&\,\fft{2\,m\,s\,\sqrt{w}}{\tilde \rho^2}\,
(c\,dt\,-\,a\,\sin^2\theta\,\fft{d\phi}{w\,\Xi})\,,\qquad A^2=A^3=0\,,\nn\\
X_1\,&=&\,H^{-\fft23},\quad X_2\,=\,X_3\,=\,H^{\fft13}\,.
\eea

   Following an analysis analogous to the one we employed in 
section \ref{abbhsec}, we find that 
the conserved energy, angular momentum and charge for 
this black hole solution are given by
\bea
E &=& \fft{\pi m[ \Xi - w(2+\Xi) + w^2 \Xi\, 
(1+\Xi)]}{4 \Xi^2 w (\Xi-w)}\,, \nn\\
J &=&\fft{\pi m a \sqrt{1-w\, \Xi}}{2 \Xi^2
\sqrt{w(w- \Xi)} }\,,\nn\\
Q &=& \fft{\pi m \sqrt{(1-w^2)(1-w\, \Xi)}}{2\Xi
\sqrt{w} (\Xi-w)}\,.\label{aejq}
\eea

\subsection{The BPS limit}

  The supersymmetry condition following from the requirement 
of a vanishing eigenvalue for the \bog matrix is now given (modulo 
equivalent sign choices) by
\be
E - g\, J - Q=0\,.
\ee
Using our expressions (\ref{aejq}) for the conserved energy, angular
momentum and charge, we find that this is BPS condition implies
\be
a^2 g^2 =\fft{1-w}{w^2}\,.
\ee
Alternatively, it can be expressed as
\be
a g = \fft{c}{s^2}\,.
\ee

    To analyse the global properties of the solution, it is
helpful to rewrite the metric as
\be
ds^2=H^{1/3} \Big(-\fft{Y\Delta_\theta \sin^2\theta\, dt^2}{(1-a^2g^2)^2 H\, 
         B_\phi} +
\fft{\rho^2 dr^2}{Y} + \fft{\rho^2 d\theta^2}{\Delta_\theta} 
   + B_\phi (d\phi + v dt)^2 +
r^2 \cos^2\theta d\psi^2\Big)\,.
\ee
This expression is valid both in the BPS limit
and in the non-extremal case.  In the supersymmetric limit there exists
a Killing vector $\ell=\del/\del t - g \del/\del \td\phi - g\del/\del\psi$
with a spinorial square root, where $\td\phi=\phi + a g^2 w c\, t$, 
and $(\td\phi,\psi)$ are asymptotically non-rotating coordinates.  From the
expression for the norm of $\ell$, which is manifestly non-positive, 
we can read off the identity
\be
-\fft{Y\Delta_\theta \sin^2\theta}{(1-a^2g^2)^2 H\, B_\phi} 
  + B_\phi (v - a g^2 w c + g)^2
+ r^2 \cos^2\theta g^2 = -H^{-1}\,.
\ee
Thus in general at the Killing horizon, where $X=0$, either $B_\phi$
or $r^2$ is negative, implying the existence of naked CTC's.  It is
straightforward to verify, unlike the previous example in section
\ref{3absec}, here naked CTC's are unavoidable when there is only one
charge and one non-vanishing rotation.  The BPS solutions therefore  all
describe naked time machines.

\section{Conclusion}

    In this paper, we have constructed new non-extremal black hole
solutions in five-dimensional $SO(6)$ gauged supergravity.  The new
solutions go beyond what has been found previously, by having unequal
values for the angular momenta in the two orthogonal 2-planes in the
transverse space.  This means that the metrics are considerably more
complicated, since they are of cohomogeneity 2, rather than the
cohomogeneity 1 of all the previously known examples.

   We have found two classes of solutions.  In the first, the two
rotation parameters are independently specifiable, as also are the
mass, and a parameter that characterises the three electric charges
carried by the gauge fields of the abelian $U(1)^3$ subgroup of the
$SO(6)$ gauge group.  Having obtained the non-extremal black hole
solutions, we then calculated the conserved angular momenta and
charges, and, by integrating the first law of thermodynamics, the
energy.  Using these results we then studied the BPS limits that give
rise to supersymmetric backgrounds.  In general, the BPS solutions
have closed timelike curves outside a Killing horizon, and hence they
describe ``naked time machines.''  However, for a special choice of
the relation between the mass and the rotation parameters, we obtain a
completely regular black hole, with neither CTC's nor singularities
outside the event horizon.  Since the two rotation parameters still
remain as free parameters, these black hole solutions provide a
continuous supersymmetric interpolation between certain previously
obtained equal-rotation solutions.  Our new solutions provide the
first examples of supersymmetric black holes in gauged supergravity in
which there are independent rotation parameters.  We also find, in
another special case, a solution describing a completely non-singular
soliton.

   We have also found a second new class of non-extremal black hole
solutions, independent of the first class, in which just one rotation
parameter is non-zero, and only one of the three $U(1)$ charges is 
non-vanishing.  We again calculated the conserved angular
momentum, charge and energy, and studied the BPS limits.  In this case, 
we find that there are no regular supersymmetric black holes or solitons,
but rather, the BPS solutions describe backgrounds with closed timelike
curves outside a Killing horizon.

\section*{Acknowledgement}

   We are grateful to Gary Gibbons for discussions.
M.C. thanks the George P. \& Cynthia W. Mitchell Institute for
Fundamental Physics for hospitality during the course
of this work.

\end{document}